\documentclass[multphys,vecphys]{svmult}

\usepackage{makeidx}     
\usepackage{graphicx}    
\usepackage{multicol}    

\makeindex             

\begin{document}

\title*{Using Multi-Band Photometry to Classify Supernovae }
\author{Dovi Poznanski, Avishay Gal-Yam, Dan Maoz\inst{1}, \\
Alexei V. Filippenko\inst{2}, Douglas C. Leonard\inst{3},
and Thomas Matheson\inst{4}}


\institute{School of Physics \& Astronomy, Tel-Aviv University, Tel-Aviv
69978, Israel
\texttt{(dovip, avishay, dani)@wise.tau.ac.il}
\and
Department of Astronomy, 601 Campbell Hall, University of
California, Berkeley, CA 94720-3411
\texttt{alex@astro.berkeley.edu}
\and
Five College Astronomy Department, University of Massachusetts, Amherst, MA 01003-9305
\texttt{leonard@corelli.astro.umass.edu}
\and
Harvard-Smithsonian Center for Astrophysics, 60 Garden Street, Cambridge, MA 02138
\texttt{tmatheson@cfa.harvard.edu}
}
%
%
\authorrunning{Poznanski et al.}
\maketitle


\section{Abstract}
\label{ABS}
Large numbers of supernovae (SNe) have been discovered in recent years, and
many more will be found in the near future. Once discovered, further study of a
SN and its possible use as an astronomical tool (e.g., as a distance estimator)
require knowledge of the SN type. Current classification methods rely almost
solely on the analysis of SN spectra to determine their type. However,
spectroscopy may not be possible or practical. 
We present a classification method for SNe
based on the comparison of their observed colors with synthetic ones,
calculated from a large database of multi-epoch optical spectra of nearby
events. 
Broadband photometry at optical wavelengths allows classification
of SNe up to $z = 0.75$, and the use of
infrared bands extends it further to $z = 2.5$. 
We demonstrate the applicability of this method, outline the observational
data required to further improve the usefulness of the method, and discuss
prospects for its use on future SN samples. Community access to the tools
developed is provided by a dedicated website.\footnotemark[1]

\footnotetext[1]{See http://wise-obs.tau.ac.il/$\sim$dovip/typing}
\section{Introduction}
Once discovered, the study of a particular supernova (SN), and its use as a tool for any 
application, is almost always based on spectroscopic verification
and classification, but this follow-up may not always be a viable option.  
First, the interesting population of very distant ($z > 1$) supernovae (SNe), 
where the mark of cosmic deceleration should be detected, 
is usually fainter than 25 mag, 
the practical limit for spectroscopy.
Second, the fast evolution of astronomical 
archives, toward the realization of the ``virtual observatory'' concept, 
is expected to produce a large number of SNe discovered in retrospective
studies, for which any follow up is, of course, impossible. 
And last, spectroscopy is impossible if large numbers (hundreds or
thousands) of SNe are detected within a relatively short time.

\begin{figure}[h]
\centering
\includegraphics[height=7.5cm]{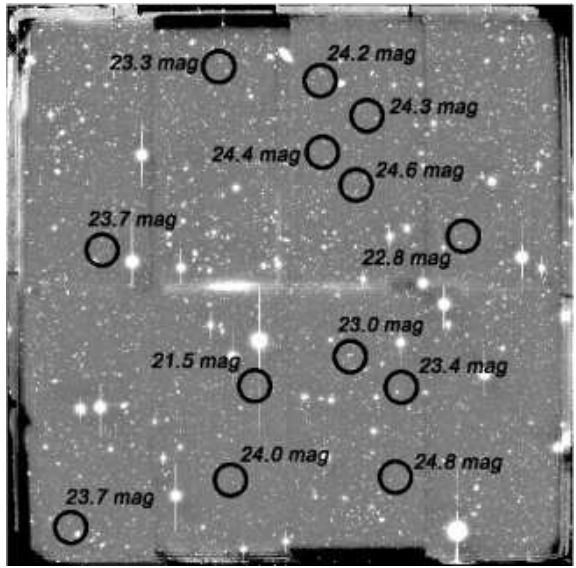}
\caption{Complete field ($36' \times 36'$) taken with the wide field mosaic camera on the 
KPNO 4m telescope. Candidate SNe with their approximate brightnesses are marked. There are
tens of still uncertain, and dimmer, candidates not marked here. }
\label{fig:mos}
\end{figure}

In an example of such occurrence,  
we have acquired with B. Jannuzi, deep (lim. mag $\sim$26) and wide (36$'\times$36$'$) 
images with the mosaic camera mounted on the KPNO 4m telescope,
on two epochs. The difference image of these two frames, reveals dozens of candidate SNe
of which only the brightest, which can be seen in Figure \ref{fig:mos}, 
can be observed spectroscopically. We have confirmation for two of these objects,
which are indeeed SNe Ia at redshifts of 0.21 and 0.67 \cite{circ}. Since these SNe are in a well studied field,
we have redshifts for most of the galaxies, and hence for the candidate SNe, and can compute
the SN redshift distribution in addition to the faint SN counts. As we will show below, a similar
dataset with photometry in more than one band would allow a separation of such an analysis
to the different SN types.

The next alternative to spectroscopy is the use of multi-color broadband 
photometry to classify SNe. 
In a recent paper \cite{P1} we have shown that in the absence of spectral 
information, much can be gleaned from the broad band colors of the SN. 
First, most SNe discovered are associated with host galaxies, so 
their redshift is either known, or can be measured long after the SN has faded 
(using spectroscopy or photometric redshifts).
Second, by comparing the colors of the SN in question to those expected for 
SNe of various types, at various ages, and at the known redshift, 
one can determine, or at least constrain, its age and type.

\section{Method}
In order to compute the expected colors of SNe of all types at a chosen $z$, 
we have compiled a large database of high signal-to-noise spectra 
of nearby SNe, mostly from Lick Observatory. 
We can then derive synthetic photometry from these spectra at any given redshift. 
The use of synthetic colors is the key that allows the tracing of the color behavior,
i.e., the time evolution of the the colors, of the various SN types, 
at arbitrary redshift. Using color-color diagrams one can then look for areas 
that are either 
populated by one type of SNe or not populated by others. Our  method allows the 
classification of SNe up to a redshift of 0.75 using optical bands, while the use 
of infrared bands increases the range up to $z=2.5$. 

\subsection{Example Diagram \& Application}
\begin{figure}[h]
\centering
\includegraphics[height=7cm]{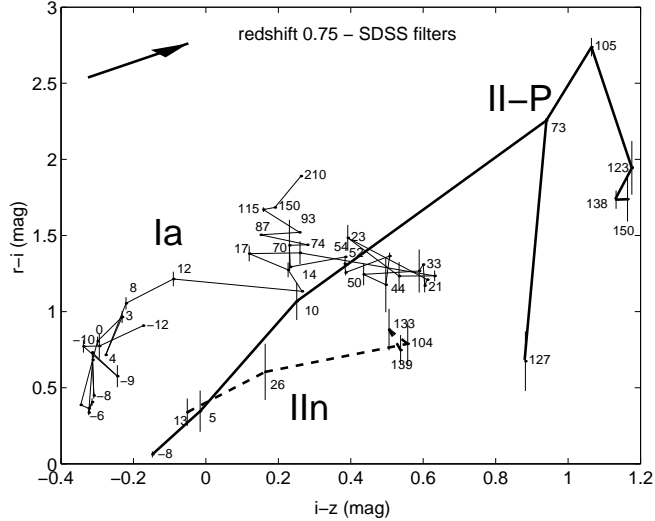}
\caption{$r-i$ vs. $i-z$ for SNe of type Ia (line), II-P (thick line), and IIn(thick-dashed line), 
at $z = 0.75$.  
The time evolution of each SN type is illustrated by
linearly connecting, in temporal order, the locations occupied by such events.
The marked ages are in days relative to $B$-band maximum. For clarity, we have
sometimes omitted the age labels of data points with similar ages and colors.
The arrow shows the reddening effect of $A_{V} = 1$ mag (in restframe $V$) 
extinction by dust in the host galaxy.}
\label{fig:riz}       
\end{figure}

Figure \ref{fig:riz} shows an example of the color paths of SNe of type Ia, II-P, and IIn
(other types are omitted for clarity), 
at $z= 0.75$ in the Sloan colors $r-i$ versus $i-z$. 
The numbers along the curves represent the time in days relative to maximum $B$-band 
brightness. One sees that most SNe before maximum light are blue, with $i-z < 0$~mag, 
but Ia SNe are significantly bluer with $i-z \sim -0.25$~mag, while being redder in
$r-i$. 
Older SNe II-P have the reddest colors of all SNe with $i-z > 1$~mag,
and IIn SNe have consistently lower $r-i$ colors than Ia SNe.
The arrow shows the reddening effect corresponding to $A_V = 1$~mag of extinction
in the host galaxy, 
assuming the Galactic reddening curve of Cardelli et al. (1989). 
One can see that the unique colors of young ($t < 12$~days) and very old ($t>100$~days) 
SNe Ia cannot be masked even by significant reddening in their host, so that 
some candidates with appropriate observed colors can be uniquely determined to be SNe Ia.
Even when the type of a SN cannot  be
uniquely determined, it may still be deduced
by supplementing the color information with other available data on the SN,
such as constraints on its brightness, and information (even if very limited)
on its variability (e.g., whether its flux is rising or declining).

Since the classification of a 
particular SN at a given redshift and with specific observed colors 
requires the generation of custom diagrams, we have built a dedicated web site 
(http://wise-obs.tau.ac.il/$\sim$dovip/typing/) that allows the astronomical 
community access to the tool have we developed.

\begin{figure}[h]
\centering
\includegraphics[height=7cm]{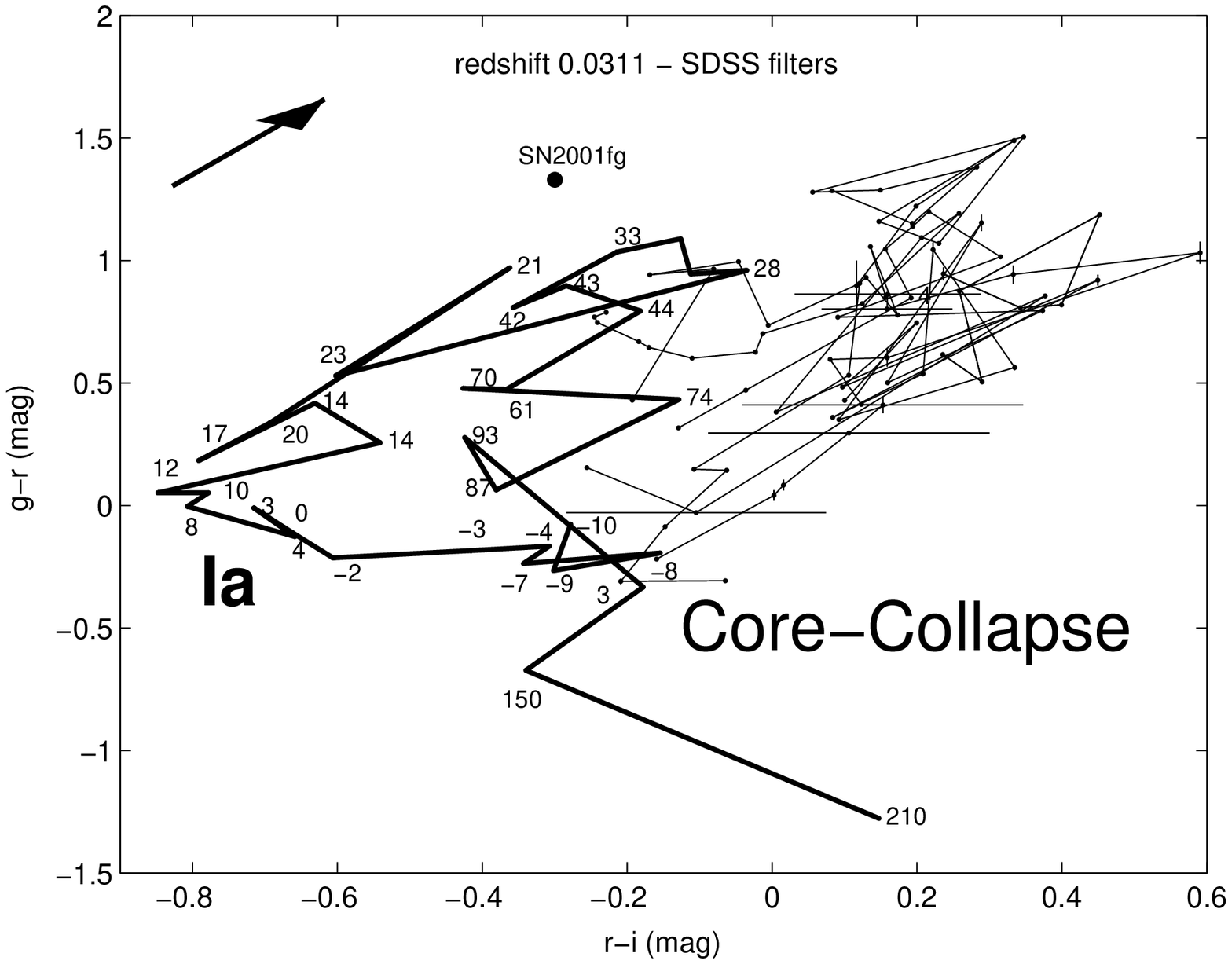}
\caption{$g-r$ vs. $r-i$ for SNe of all types at $z = 0.0311$. SNe Ia tracks are thick lines, 
and core-collapse SNe tracks are thin lines.
The observed colors of SN 2001fg are indicated by the filled circle. The colors
suggest a SN~Ia about 1 month past maximum brightness, as confirmed by spectra
of this event.}
\label{fig:sdss}      
\end{figure}

An example of such an application is shown in Figure \ref{fig:sdss}. 
SN 2001fg was one of the first SNe reported from SDSS data. This event was discovered
on 15 October 2001 UT \cite{vanden}  with $g = 19.20$, $r =
17.87$, and $i = 18.17$ mag. Inspecting the color-color diagram calculated for
the appropriate redshift ($z = 0.0311$, see below), one can see (Figure \ref{fig:sdss}) that
the type and approximate age of the SN candidate can be deduced.  The diagram
clearly indicates this is a SN~Ia, about one month old.  Followup spectra \cite{fili}
 using the Keck II 10-m telescope reveal that the
object is indeed a SN~Ia, at $z = 0.0311$. The spectrum is similar to those of
SNe~Ia about two months past maximum brightness.  This age, at the time of the
spectral observation (18 November UT), implies an age around one month past
maximum brightness at discovery, confirming our diagnosis. A fully ``blind''
application would have, of course, required an independent photometric or
spectroscopic redshift for the SN host galaxy.

\section{Conclusions and Future Prospects}

We have shown that SNe can be classified using broad band colors, so that spectroscopy
is not a prerequisite. Such an approach will enable the analysis of 
larger number of SNe, as expected to be discovered in the coming years, 
but with limited information. This is important especially at high $z$ 
where spectroscopy becomes almost impossible.

In order to better sample the evolution of the covered SN types, 
we plan to enrich the spectral database used. 
This will smooth the current scatter
seen, e.g., in Figure \ref{fig:riz}, which is largely due to variations
 between individual objects. 
We also intend to incorporate spectra of 1998bw-like Ic SNe, also 
called hypernovae \cite{ham03}, associated with gamma-ray bursts. 
Finally, in order to enable classification
of higher redshift SNe observed in optical bands, UV spectra, of which only a handful exist 
\cite{pana}, will be added to the database.

%
%

%
%

\printindex
\end{document}